\documentclass[aps,pra,twocolumn,superscriptaddress,notitlepage,nofootinbib,accepted=2018-10-25]{quantumarticle}

\usepackage[utf8]{inputenc}
\usepackage[english]{babel}

\usepackage{amsmath,amsfonts,amssymb}
\usepackage{amsthm}
\usepackage{amsbsy}

\usepackage{mathtools}
\usepackage{latexsym}
\usepackage{relsize}
\usepackage{mathrsfs}

\usepackage{euscript}
\usepackage{graphicx}

\usepackage{bbm}
\usepackage{bm}
\usepackage{epsfig}
\usepackage{epstopdf}
\usepackage{dsfont}

\usepackage[colorlinks]{hyperref}
\usepackage[figure,table]{hypcap}
\usepackage{enumitem}
\usepackage{geometry}
\hypersetup{
	bookmarksnumbered,
	pdfstartview={FitH},
	citecolor={darkgreen},
	linkcolor={darkblue},
 linktoc={page},
	urlcolor={darkblue},
	pdfpagemode={UseOutlines}}

\usepackage{color}
\definecolor{darkgreen}{RGB}{40,130,40}
\definecolor{darkblue}{RGB}{0,0,190}
\definecolor{darkred}{RGB}{238,0,0}
\definecolor{purple}{RGB}{128,0,255}
\usepackage{soul}
\usepackage{float}
\usepackage{algorithm}

\newtheorem{theo}{Theorem}
\newtheorem{prop}{Proposition}

\newtheorem{coro}{Corollary}

\newcommand{\alg}[1]{\hyperref[alg:#1]{Alg.~\ref*{alg:#1}}}
\newcommand{\app}[1]{\hyperref[app:#1]{App.~\ref*{app:#1}}}
\newcommand{\thm}[1]{\hyperref[thm:#1]{Thm.~\ref*{thm:#1}}}
\newcommand{\prp}[1]{\hyperref[prp:#1]{Prop.~\ref*{prp:#1}}}
\newcommand{\fig}[1]{\hyperref[Fig#1]{Fig.~\ref*{Fig#1}}}

\def\one{\mathbb{I}}
\def\hil{\mathcal{H}}

\def\M{\mathcal{M}}

\def\FPAA{\textsf{FPAA}}
\def\PFS{\textsf{PrepareFromSamples}}
\def\PFU{\textsf{PrepareFromUniform}}
\def\O{\mathcal{O}}
\def\OO{\widetilde{\mathcal{O}}}
\def\ppi{{\boldsymbol{\pi}}}
\def\ppim{\boldsymbol{\pi}^\mathcal{M}}
\def\err{\varepsilon}

\newcommand{\ket}[1]{{\left\vert{\,#1\,}\right\rangle}}
\newcommand{\braket}[2]{{\left\langle{\,#1\,}\left\vert{\,#2\,}\right.\right\rangle}}
\newcommand{\ketbra}[1]{\left|\,#1\,\right\rangle\!\left\langle\,#1\,\right|}
\newcommand{\norm}[1]{\left\lVert{#1}\right\rVert}
\newcommand{\II}{\mathpalette\raiseII\relax}
\newcommand{\raiseII}[2]{\raisebox{-.4pt}{$#1\mathds{I}$}}

\begin{document}

\title{Faster quantum mixing for slowly evolving sequences of Markov chains}

\author{Davide Orsucci}
\email{davide.orsucci@uibk.ac.at}
\affiliation{Institute for Theoretical Physics, University of Innsbruck, Technikerstra{\ss}e 21a, 6020 Innsbruck, Austria}
\orcid{0000-0003-3087-8757}

\author{Hans J. Briegel}
\email{hans.briegel@uibk.ac.at}
\affiliation{Institute for Theoretical Physics, University of Innsbruck, Technikerstra{\ss}e 21a, 6020 Innsbruck, Austria}
\affiliation{Department of Philosophy, University of Konstanz, Fach 17, 78457 Konstanz, Germany}
\orcid{0000-0002-9065-1565}

\author{Vedran Dunjko}
\email{v.dunjko@liacs.leidenuniv.nl}
\affiliation{Institute for Theoretical Physics, University of Innsbruck, Technikerstra{\ss}e 21a, 6020 Innsbruck, Austria}
\affiliation{Max-Planck-Institut f\"ur Quantenoptik, Hans-Kopfermann-Str.\ 1, 85748 Garching, Germany}
\affiliation{LIACS, Leiden University, Niels Bohrweg 1, 2333 CA Leiden, The Netherlands}
\orcid{0000-0002-2632-7955}

\begin{abstract}
Markov chain methods are remarkably successful in computational physics, machine learning, and combinatorial optimization. The cost of such methods often reduces to the mixing time, \textit{i.e.},\ the time required to reach the steady state of the Markov chain, which scales as $\delta^{-1}$, the inverse of the spectral gap. It has long been conjectured that quantum computers offer nearly generic quadratic improvements for mixing problems. However, except in special cases, quantum algorithms achieve a run-time of $\O(\sqrt{\delta^{-1}} \sqrt{N})$, which introduces a costly dependence on the Markov chain size $N,$ not present in the classical case. Here, we re-address the problem of mixing of Markov chains when these form a slowly evolving sequence. This setting is akin to the simulated annealing setting and is commonly encountered in physics, material sciences and machine learning. We provide a quantum memory-efficient algorithm with a run-time of $\O(\sqrt{\delta^{-1}} \sqrt[4]{N})$, neglecting logarithmic terms, which is an important improvement for large state spaces. Moreover, our algorithms output quantum encodings of distributions, which has advantages over classical outputs. Finally, we discuss the run-time bounds of mixing algorithms and show that, under certain assumptions, our algorithms are optimal.
\end{abstract}

\maketitle

\section{Introduction}
\label{sec:Intro}

Markov chains (MCs) are central in computational approaches to physics~\cite{1999_Newman}, in computer science~\cite{1993_Sinclair}, and machine learning~\cite{1957_Bellman}, and they form the crux of the ubiquitous Markov Chain Monte Carlo methods~\cite{1995_Gilks}. In MC-based approaches the underlying objective is to produce samples from the steady state, \textit{i.e.},\ the stationary distribution of a given MC. The MC is constructed so that this distribution encodes the solution of the problem at hand. The solution can then be reached by ``mixing'', \textit{i.e.},\ by applying the MC transitions many times.  For some problems, mixing processes constitute the fastest known classical solving algorithms, and play a vital role, \textit{e.g.}, in the Metropolis-Hastings methods~\cite{1970_Hastings}, periodic Gibbs sampling~\cite{1987_Geman}, and Glauber dynamics~\cite{1999_Martinelli}.

The fundamental parameter governing the time complexity of MC-based algorithms is thus the \emph{mixing time}, that is, the number of steps required to attain stationarity. In most applications the MC is ergodic, \textit{i.e.},\ has a unique stationary distribution, and time-reversible, \textit{i.e.},\ satisfies detailed balance~\cite{1998_Norris,2017_Levin}. The mixing time is tightly related to the spectral gap $\delta$ of the MC\footnote{The spectral gap is defined with $\delta = 1 - |\lambda_2|,$ where  $\lambda_2$ is the second largest eigenvalue  (in absolute value) of the transition matrix of the time-reversible Markov chain.} and is bounded by $\Omega{(\delta^{-1})}$~\cite{1997_Aldous}.

\begin{figure*}
\includegraphics[width=\textwidth]{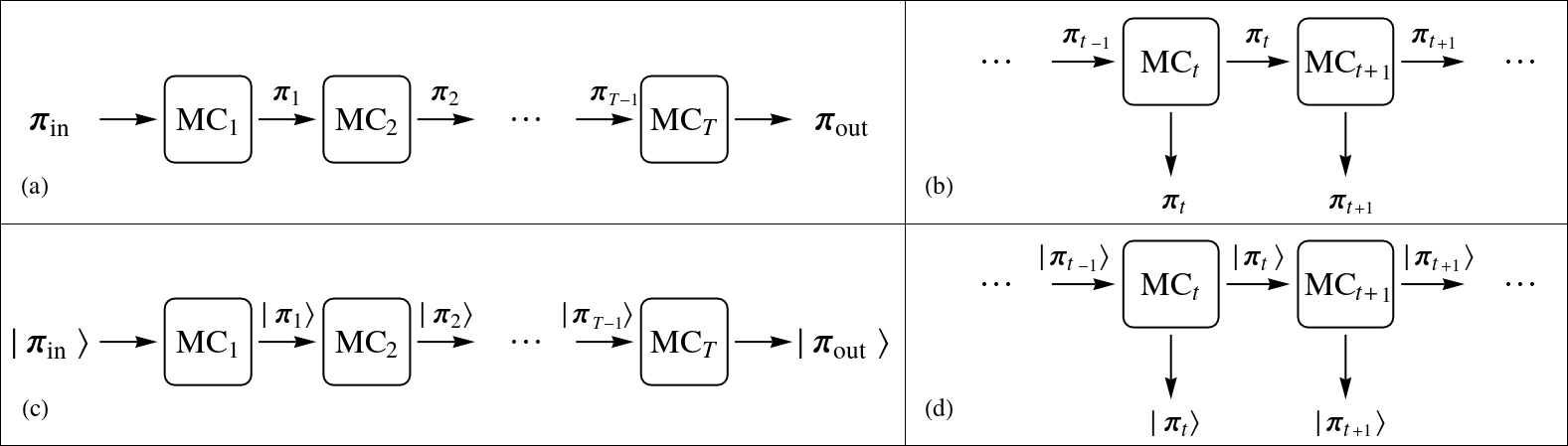}
\caption{ Schematic depiction of the scenarios of sequences of slowly evolving MCs. Panels (a,b) and (c,d) depict classical and quantum sampling tasks, respectively. Panels (a,c) and (b,d) respectively delineate finite and continuing sequences, the latter having step-wise outputs. This work is predominantly concerned with (b,d). Although panel (d) allows quantum states to be carried from one time-step to another, our algorithm actually works by forwarding just classical information, without sacrificing efficiency. That is, no quantum memory from one time-step to the next is required.}
\label{Fig1}
\end{figure*}

Oftentimes direct mixing can be computationally prohibitive and thus heuristic methods, such as \emph{simulated annealing}~\cite{1983_Kirk,1987_Laar}, are employed. Here one constructs a sequence of Markov chains which, for instance, encode the Gibbs (thermal) distributions at gradually decreasing values of the temperature, where the target distribution is specified by the final MC, \textit{i.e.}, the final temperature. Intuitively, this process increases efficiency by avoiding local minima, although the performance is typically not guaranteed. In simulated annealing, the neighbouring chains in the sequence are similar, in other words, the sequence is \emph{slowly evolving}.

The emergence of quantum computation offers a new possibility to utilize quantum effects to achieve guaranteed mixing more rapidly. In particular, it has been conjectured that run-times of $\OO\big(\sqrt{\delta^{-1}}\big)$\footnote{For expressing run-times we adopt the soft-$\O$ notation ($\OO$), an extension of the $\O$ notation where polylogarithmic multiplicative factors are neglected.} should be possible~\cite{2007_Richter} for the mixing problem. Such quadratic speed-ups have been demonstrated for various special cases of MCs~\cite{2000_Nayak,2001_Ambainis,2001_Aharonov,2007_Richter, 2007_Richter_NJP,2015_Dunjko2}, mostly relying on quantum walk~\cite{2003_Kempe, 2011_Reitzner} approaches. Quantum walks have also been utilized to speed-up simulated annealing~\cite{2008_Wocjan, 2008_Somma_PRL, 2009_Wocjan}, which often leads to the best run-times in practice. However, considering provable results for guaranteed mixing of general Markov chains, the best quantum algorithms achieve $\OO\big(\sqrt{\delta^{-1}} \sqrt{N}\big),$ which falls short of the conjectured quadratic speed-up, as it introduces the dependence on the system size $N$. Avoiding the $\O\big(\sqrt{N}\big)$ dependence seems to be challenging, which further motivates investigating the settings with relaxed constraints, \textit{e.g.}, by restricting the MC family~\cite{2004_Childs,2007_Richter,2015_Dunjko2}.

In this work, we obtain improved $\OO\big(\sqrt{\delta^{-1}} \sqrt[4]{N}\big)$ run-times for mixing problems not by restricting the Markov chain family, but rather by relying on additional context. In particular, we consider the settings where we are tasked to sequentially produce \emph{independent samples} from a sequence of slowly evolving Markov chains. This setting is natural in statistical and quantum physics, \textit{e.g.},\ when studying phase boundaries, which requires many independent samples from near-by points in the parameter space~\cite{2010_Nishimori}. Another motivation for studying this setting is in the context of machine learning (ML), appearing both in reinforcement learning~\cite{1998_Sutton} and in the training of generative models~\cite{2016_Bishop}, as we discuss later in the paper.

Our setting is similar to simulated annealing in that it considers a sequence of pair-wise similar Markov-chains, and indeed our methods are similar to those in \cite{2008_Wocjan, 2008_Somma_PRL}. However our setting brings about a key distinction: in annealing the goal is to produce a sample from final MC, and the intermediary chains have only an auxiliary role; in our case the goal is to produce independent samples from each MC in the sequence. Further, in principle the sequence can be exponentially large, or having a length which is not a priori specified. This is schematically illustrated in \fig{1}.

\section{Problem and methods}
\label{sec:problem}

We now specify the setting more precisely and introduce the required notation. We consider finite-space MCs, of size $N$, where a distribution over the space is specified by a vector $\ppi := (\pi(1), \pi(2), \ldots, \pi(N))^T$ of non-negative entries summing to one. The problem of sampling from this distribution corresponds to producing a single element $x \in \{1,2,\ldots, N\}$ according to $\ppi$. Due to the methods used, our algorithms will actually produce a quantum (or coherent) sample, \textit{i.e.},\ the quantum state $\ket{\ppi} := \sum_x \sqrt{\pi(x)} \, \ket{x}$, which is called the \emph{coherent encoding} of $\ppi$. 
As we elaborate later, coherent encodings have substantial advantages over classical samples, although they are in general computationally more difficult to prepare.

In abstract terms our sampling problem can be formulated as follows. We consider an infinite sequence of ergodic time-reversible Markov chains $\{ \textup{MC}_t \}_{t=1}^{\infty}$. In our approach, we will at each time-step $t$ generate a coherent sample, that is the state $\ket{\ppi_t}$, corresponding to the stationary distribution $\ppi_t$ of MC$_t$. The measurement of $\ket{\ppi_t}$ in the computational basis yields a classical sample from $\ppi_t$, thus preparation of $\ket{\ppi_t}$ allows for both classical and quantum sampling. We say that the sequence of MCs is slowly evolving if the stationary distributions of consecutive MCs in the sequence are sufficiently close, specifically, if at every time step $t$ we have $| \braket{\ppi_{t+1}}{\ppi_{t}}|^2  \geq \eta$ for some constant $0 < \eta < 1$.

Our techniques rely on Szegedy-type quantum walks~\cite{2004_Szegedy} to perform the above sequential sampling task with a $\OO(\sqrt{\delta^{-1}}\sqrt[4]{N})$ time complexity in the context of slowly evolving MCs. We thus briefly introduce the properties of the Szegedy constructions for the convenience of the reader and provide in \app{MC} more background on MC theory.

\subsection{Szegedy quantum walk}
\label{sec:szegedy}

Each MC is specified by a stochastic matrix $P$ which specifies the transition probabilities in a single step of the chain. For a given transition matrix $P$ of a ergodic time-reversible MC one can construct the corresponding Szegedy quantum walk operator $W(P)$; this is a unitary operator having the crucial property that $\ket{\ppi}$ is the unique $+1$-eigenstate of $W(P)$, with all other eigenstates of $W(P)$ having an eigenphase which is at least quadratically larger than the spectral gap $\delta$ of $P$. In other words, if $\ket{\theta}$ is such that $W(P)\ket{\theta} = \textup{e}^{i \theta} \ket{\theta}$, then $|\theta| \in \O(\sqrt{\delta})$, see \app{Szegedy} for details and the construction.

These properties allow us to realize useful quantum subroutines with a run-time which is quadratically smaller than the classical mixing time. Namely, the Szegedy walk operator can be used in conjunction with the phase detection algorithm~\cite{2011_Magniez_SIAM}, a simple variant of phase estimation~\cite{1996_Kitaev}, to (approximately) distinguish $\ket{\ppi}$ from all other eigenstates of $W(P)$. The run-time is in $\OO\big( \sqrt{\delta^{-1}} \big)$ and has only logarithmic dependence on the approximation error\footnote{All algorithms we consider are approximate, but the dependence on the target error $\err$ is at most $\log^2(\err^{-1})$ and thus always ignored in the $\OO$ notation.}~\cite{1996_Kitaev, 2011_Magniez_SIAM, 2013_Svore, 2016_Wiebe}. In turn, the capacity to identify $\ket{\ppi}$ can be leveraged to implement an approximate projective measurement onto $\ketbra{\ppi}$, by measuring whether the quantum register containing the phase estimate is zero. Similarly, applying a Pauli-$Z$ rotation onto the qubit that specifies whether the phase value is zero we obtain an approximation of the reflection operator $\textup{R}(\ppi) := \II - 2 \ketbra{\ppi}$.

\subsection{Amplitude amplification}
\label{sec:amplitude_amplification}

This brings us to our key subroutine. Using the reflection $\textup{R}(\ppi)$ we use amplitude amplification~\cite{1996_Grover,2000_Brassard} to rotate an initial state $\ket{\psi_{in}}$ to an approximation of $\ket{\ppi}$ in time $\OO \big( \sqrt{\gamma^{-1}} \sqrt{\delta^{-1}} \big)$, where $| \braket{\psi_{in}}{\ppi} |^2 \geq \gamma$. Equivalently, we can also use the fixed-point amplitude amplification algorithm\footnote{The fixed-point property means that the output state converges to the ideal target state for increasing run-times~\cite{2005_Grover}.} of Yoder \emph{et al.}~\cite{2014_Yoder}, an algorithm that has the same quadratic speed-up as standard amplitude amplification. We refer the reader to \app{subroutines} for further details on fixed-point amplitude amplification and to \app{imperfect_ref} for an analysis of how the runtime depends on the target precision.

With these key subroutines defined we can explain a straightforward algorithm that allows to prepare coherent encodings of $\ket{\ppi}$ with $\O\big(\sqrt{\delta^{-1}} \sqrt{N}\big)$ run-time. One simply utilizes amplitude amplification to rotate the uniform superposition $\ket{\textbf{u}} := \frac{1}{\sqrt{N}} \sum_{x} \ket{x}$ to the target state $\ket{\ppi}$. Since the target states are encodings of probability distributions, all amplitudes are real and non-negative and, thus, the fidelity always satisfies $\gamma > 1/N$. This bound is attained by distributions approaching a Kronecker delta.

We remark that amplitude amplification requires the ability to reflect both around the source state $\ket{\textbf{u}}$ and the target state, \emph{i.e.}, to implement both $\textup{R}(\ppi)$ and $\textup{R}(\textbf{u}) := \II - 2 \ketbra{\textbf{u}}$. In this work we restrict our attention to cases in which the preparation of $\ket{\textbf{u}}$ is efficient and, therefore, also $\textup{R}(\textbf{u})$ can be easily implemented. Notice that preparing the uniform distribution is not always simple~\cite{2007_Richter_NJP}, \emph{e.g.}, this happens when it is computationally difficult to decide if an element $x \in \mathds{N}$ belongs to the space of the MC\footnote{As a concrete example, consider the following MC inspired by the Graph Isomorphism problem: the space of the MC consists of all graphs that can be obtained from an initial graph via permutation of the vertices of a given initial graph; and a transition in the MC is obtained by randomly selecting two vertices of a graph and swapping them. Then, deciding if a graph belongs to the space is equivalent to solving the Graph Isomorphism problem.}. However, assuming that the space of the MC is $\{1,\ldots, N\}$ the uniform distribution can be prepared with quantum circuits of depth $\O(\log (N))$~\cite{2002_Grover}; this is a case that finds application to quantum machine learning problems~\cite{2014_Paparo}. Even more simply, when $N = 2^n$ the uniform distribution is obtained from $\ket{0}$ via the Hadamard transform, that is, $\ket{\textbf{u}} = H^{\otimes n} \ket{0}$. Consequently, our methods can be applied to spin $1/2$ systems, where the preparation of the uniform superposition of all the configuration of $n$ spins is trivial, yet producing Gibbs distributions at low temperature for certain classical Hamiltonians is \textsf{NP}-hard~\cite{2010_Sly}.

\section{Preparation from uniform distribution and from samples}
\label{sec:unif_and_sample}

To speed-up the basic algorithm described in the previous section, the idea is to eliminate the worst-case preparation scenario. Specifically, in the case when the distribution is highly clumped, one should attempt the preparation from an element having high probability in the target distribution $\ppi$. However, we still have to choose the candidate element to start from, which alone would lead to a $\Omega\big(\sqrt{N}\big)$ run-time (by the optimality of Grover's search \cite{1998_Boyer}). We will first show that this issue can be resolved when one has access beforehand to a few classical samples from the target distribution. This seems to require that the solution we are looking for are already provided as input. But we will utilize the slowly evolving context to ensure such samples are available, and therefore samples for the subsequent step can be prepared without the necessity of back-tracking in the sequence.

To utilize these ideas we first show how to prepare the coherent encoding $\ket{\ppi}$ by choosing a suitable initial state $\ket{\psi_{in}}$ and then amplitude amplify $\ket{\psi_{in}}$ to obtain $\ket{\ppi}$. Specifically, the initial state is either the uniform distribution, $\ket{\psi_{in}} \equiv \ket{\textbf{u}}$, or a classical sample $x_j$ which is taken from a small set of classical samples $\vec{x} = \{x_1, \dots, x_c\}$ that are available beforehand, $\ket{\psi_{in}} \equiv \ket{x_j}$. We will call these subroutines $\PFU$ and $\PFS$, respectively, and simply $\textsf{Prepare}$ whenever the distinction is not relevant.

As mentioned, $\PFU$ is efficient in the extreme case where $\ppi$ is very close to being uniform, while the procedure can require up to $\O(\sqrt{N})$ operations in the opposite extreme case where $\ppi$ has support over only one element $x$; this last case corresponds, in fact, to a standard Grover search for the element $x$. However, when most of the ``weight'' (probability) of $\ppi$ is concentrated on a few elements (which need not be nearby) these must have a large overlap with $\ket{\ppi}$, which is a sufficient condition to efficiently perform amplitude amplification; that is, running a search algorithm in reverse (un-searching) from one of these elements allows for a fast re-preparation of $\ket{\ppi}$~\cite{2004_Childs}. Then, a classical sample drawn from $\ppi$ will probably come from elements having large ``weight'' and thus the $\PFS$ subroutine will be efficient. The main idea of our algorithms is to discover which of the two $\textsf{Prepare}$ algorithms is the most efficient and then use it for state preparation. The worst regime is for distributions which are neither too uniform nor too clumped, where both algorithms have a $\O\big(\sqrt[4]{N}\big)$ time complexity.

Before continuing with the complete description of the $\textsf{Prepare}$ subroutines, we remark that we use amplitude amplification starting from $\ket{\psi_{in}}$ to produce $\ket{\ppi}$ and thus we require the ability to perform reflections both around $\ket{\psi_{in}}$ and around $\ket{\ppi}$. Both choices for the initial state can be prepared efficiently: $\ket{x_j}$ is simply a classical state, while $\ket{\textbf{u}}$ can be prepared easily when the MC space is explicitly known. Thus, also reflectors around them can be efficiently implemented. A reflection around $\ket{\ppi}$ is instead approximated with $\OO(\sqrt{\delta^{-1}})$ operations using Szegedy operator. Therefore, the total gate cost of $\textsf{Prepare}$ is $\OO\big( \sqrt{\gamma^{-1}} \sqrt{\delta^{-1}} \, \big)$ where $|\braket{\psi_{in}}{\ppi}|^2 \geq \gamma$. We are then left with the task of estimating this lower bound $\gamma$.

\subsection{Preparing from uniform}
\label{sec:unif}

This subroutine is the straightforward algorithm we mentioned earlier. Suppose for the moment that the value of $|\braket{\textbf{u}}{\ppi}|$ is known. Then we can amplitude amplify $\ket{\textbf{u}}$ to $\ket{\ppi}$, operation having a gate cost which is proportional to
\begin{align}
\label{eq:bound1}
	\big|\braket{\textbf{u}}{\ppi}\big|^{-1} 
	\; = \; 
	\frac{\sqrt{N}}{f(\ppi)} \;,
\end{align}
in which we have introduced the notation:
\begin{align}
	f(\ppi) 
	\; := \;
	\sum_{x=1}^N \sqrt{\pi(x)} \;.
\end{align}
By norm inequalities we get $1 \leq f(\ppi) \leq \sqrt{N}$, where the lower and upper bounds are saturated by a Kronecker delta and the uniform distribution, respectively.

If the value of $|\braket{\textbf{u}}{\ppi}|$ is not known we proceed as follows. We arbitrarily choose a value $\chi' \equiv \sqrt{N} / \chi$ as a tentative estimate of $\sqrt{N} / f(\ppi)$ so that amplitude amplification produces an approximation of $\ket{\ppi}$ when $|\braket{\textbf{u}}{\ppi}| \geq \chi'^{-1}$ holds\footnote{This can be easily enforced using the fixed-point version of amplitude amplification.}. However, we do not know if the initial overlap is large enough and thus preparation of $\ket{\ppi}$ is not guaranteed. To amend this, we subsequently apply to the output of amplitude amplification a projective measurement onto $\ketbra{\ppi}$ (or onto the orthogonal complement) which, if successful, heralds the correct preparation of $\ket{\ppi}$. As we mentioned, this projective measurement can be implemented with Szegedy quantum walks with run-time $\OO\big(\sqrt{\delta^{-1}}\big)$ and therefore its run-time is independent from the initial overlap $|\braket{\textbf{u}}{\ppi}|$. 
This constitutes a \emph{heralded preparation} of $\ket{\ppi}$ from $\ket{\textbf{u}}$.

$\PFU$ is summarized in \alg{from_unif}, and in \app{fail_prob} further details and error analysis are provided. The run-time for preparing $c$ copies is in $\OO\big( c \, \chi'  \sqrt{\delta^{-1}} \big)$ with an exponentially decaying failure probability when $\chi' \geq \sqrt{N} / f(\ppi)$.

\begin{algorithm}[H]
\caption{$\PFU$}
\label{alg:from_unif}
\textbf{Output:} a bit signalling success; in case of success, $c$ quantum samples (\textit{i.e.}, $c$ copies of the state $\ket{\ppi}$).\vspace{2mm}\\
\textbf{Input:} quantum access to the transition matrix $P$; $c$, the number of copies to be produced; $\chi' \equiv \sqrt{N}/\chi$, a (tentative) estimate of $\sqrt{N}/f(\ppi) = |\braket{\textbf{u}}{\ppi}|^{-1}$. 
\vspace{2mm}\\
\textbf{Algorithm:}
\begin{enumerate}
\item For $j = 1, \ldots, 2c $:
\begin{itemize}[leftmargin=3mm]
\item[] Run a heralded preparation of $\ket{\ppi}$ from $\ket{\textbf{u}}$ as described in the main text, assuming a initial overlap larger than $1/\chi'$.
\end{itemize}
\item If at least $c$ successful preparations have been heralded in step~1., output a bit signalling success, together with the quantum states obtained in $c$ successful runs of heralded preparation of $\ket{\ppi}$. Else, return a bit signalling failure.
\end{enumerate}
\end{algorithm}

\subsection{Preparing from samples}
\label{sec:sample}

The second subroutine we will utilize, named $\PFS$, requires extra inputs, namely a set of $c$ samples $\vec{x} = \{x_1, \ldots, x_c\}$ from the desired target distribution $\ppi$, and is based on amplitude amplification of $\ket{x_j}$ to $\ket{\ppi}$. Later we will show how these samples can be efficiently obtained in a slowly evolving sequence.

The run-time of amplitude amplification scales as $|\braket{x_j}{\ppi}|^{-1} = 1/\sqrt{\pi(x_j)}$ assuming, for the moment being, that the value of $|\braket{x_j}{\ppi}|$ is known. We thus introduce a random variable $X$ distributed according to $\ppi$, \textit{i.e.}, $X$ takes a value $x$ with probability $\pi(x)$. The run-time of amplitude amplification is also a random variable, proportional to $|\braket{X}{\ppi}|^{-1} = \pi^{-1/2}(X)$. The average run-time scales as
\begin{align}
\label{eq:bound2}
	\mathbb{E}_{\ppi} [\,\pi^{-1/2}(X)\,] 
	& \; = \;
	\sum_{x=1}^N \pi(x) \ \pi^{-1/2}(x)
	\; = \; 
	f(\ppi)\;.
\end{align}

Note that we have bounded only the \emph{expected} run-time of our algorithm and not of a specific instance of the algorithm, \textit{i.e.},\ for a particular choice of $x_j$. The sampling procedure could return in fact a sample $x_j$ for which the run-time factor [$\pi^{-1/2}(x_j)$] is much larger than its average value.

However, this can be prevented by using a few samples, since the run-time when starting from a randomly sampled element $x_j$ is, with constant probability, close to the average run-time. To formalize this we use Markov's inequality:
\begin{align}
\label{eq:Markov}
	\Pr\{\ \pi^{-1/2}(X) \geq a \ \mathbb{E}[\ \pi^{-1/2}(X)\ ]\ \} 
	\; \leq \; 
	\frac{1}{a} \;.
\end{align}
We then consider the case $a=2$ and proceed similarly as we did for $\PFU$. Namely, we guess an estimate $\chi$ for $f(\ppi)$ and amplitude amplify $\ket{x_j}$ to $\ket{\ppi}$ assuming that $|\braket{x_j}{\ppi}| \geq 1/(2\chi)$ holds (that is, we use $\O(2 \chi)$ reflections) and then repeat for all samples in $\vec{x}$. Subsequently we apply a projective measurement onto $\ketbra{\ppi}$ to herald the successful preparation of $\ket{\ppi}$.

Suppose that $\chi \geq f(\ppi)$. Since the $c$ samples are independent, the probability that $\PFS$ fails 
for all the samples in $\vec{x}$ is then exponentially small in $c$ (for instance, $\PFS$ can fail if $|\braket{x_j}{\ppi}|^{-1} > 2 f(\ppi)$ holds for all $x_j$). A formal specification of $\PFS$ for preparing $c$ new samples is given in \alg{from_samples} and has run-time in $\OO\big( c^2 \, \chi \, \sqrt{\delta^{-1}} \big)$. The failure probability again goes down exponentially  if $\chi \geq f(\ppi)$. See \app{fail_prob} for details and error analysis.

\begin{algorithm}[H]
\caption{$\PFS$}
\label{alg:from_samples}
\textbf{Output:} a bit signalling success; in case of success, $c$ quantum samples (\textit{i.e.}, $c$ copies of the state $\ket{\ppi}$).\vspace{2mm}\\
\textbf{Input:} quantum access to the transition matrix $P$; $\vec{x} = \{x_1, \ldots, x_c\}$, a set of $c$ classical samples approximately drawn from $\ppi$ (correspondingly, we require to produce $c$ new copies of $\ket{\ppi}$); $\chi$, a  (tentative) estimate of $f(\ppi) = \mathbb{E}[\, \pi^{-1/2}(X) \,]$. \vspace{2mm}\\
\textbf{Algorithm:}
\begin{enumerate}
\item For $j = 1, \ldots, c$: 
\begin{itemize}[leftmargin=3mm]
\item[] For $2c$ times: run a heralded preparation of $\ket{\ppi}$ from $\ket{\textbf{u}}$ as described in the main text, assuming a initial overlap larger than $1/(2 \chi)$.
\end{itemize}
\item If at least $c$ successful preparations have been heralded in step~1., output a bit signalling success, together with the quantum states coming from $c$ successful runs of heralded preparation of $\ket{\ppi}$. Else, return a bit signalling failure.
\end{enumerate}
\end{algorithm}

\subsection{Combined algorithm}
\label{sec:combined}

Now we will put together the two $\textsf{Prepare}$ subroutines in a single combined algorithm. In the case in which the value $f(\ppi)$ is known one simply runs whichever of the two $\textsf{Prepare}$ algorithms is faster.

Summarizing Eq.~\eqref{eq:bound1} and Eq.~\eqref{eq:bound2} the run-time is then in $\OO \big( c^2 \, C(\ppi) \, \sqrt{\delta^{-1}} \big)$, where:
\begin{align}
\label{eq:bound_comb}
	C(\ppi) 
	\; := \;
	\min 
	\bigg\lbrace \
	\frac{\sqrt{N}}{f(\ppi)} \ , \
	f(\ppi)
	\ \bigg\rbrace \; \leq \sqrt[4]{N} \;. 
\end{align}

To deal with the situation when $f(\ppi)$ is not known, we modify the algorithm in a manner similar to how Grover's search is adapted to work without an estimate on the number of marked numbers \cite{1998_Boyer}. Essentially, one runs both preparation algorithms one after another, starting from $\chi = 1$ and $\chi'=1$ for $\PFS$ and $\PFU$, respectively; then, in each iteration the values of $\chi$ and $\chi'$ are set to twice larger values, terminating either when $c$ copies of the coherent encoding are produced or when both $\chi$ and $\chi'$ exceed $2 \sqrt[4]{N}$. Then the total number of reflections required scales as $C(\ppi)$, which is $\sqrt[4]{N}$ in the worst case, and the global failure probability is, again, exponentially small.

\section{Application in the context of slowly evolving sequences}
\label{sec:application}

The algorithm given in the preceding paragraphs can be implemented also for a stand-alone MC, \textit{i.e.}, for a chain not coming from a slowly evolving sequence. However, it comes with the unrealistic requirement that $c$ samples drawn from the stationary distribution of $\ppi$ are available beforehand: it seems that, paradoxically, the output of the algorithm is also required as input. Nonetheless, the result is non-trivial even for stand-alone MCs because of the following two observations. First, the initial classical samples can be re-used to prepare multiple coherent copies and this, in turn, allows us to prepare an arbitrary number of fresh independent samples, given only a small number ($c$) of seed examples. Second, our algorithm outputs a coherent encoding of the stationary distribution, allowing quantum information post-processing to be applied.

Going back to our primary objective, we now show how these initial samples can be made available in the context of slowly evolving sequences.

We proceed inductively. We suppose that at time step $t$ we have at hand $c$ samples from $\ppi_t$. This allows us to produce $c$ copies of $\ket{\ppi_t}$ in time $\OO \big( c^2 \, C(\ppi_t) \, \sqrt{\smash[b]{\delta_{t}^{-1}}} \big)$. Next, using a Szegedy quantum walk we can implement reflections both around $\ket{\ppi_t}$ and around $\ket{\ppi_{t+1}}$. This allows us to use amplitude amplification (or its fixed-point variant) to approximately map each of the $c$ copies of $\ket{\ppi_t}$ to a copy of $\ket{\ppi_{t+1}}$. In turn, these copies of $\ket{\ppi_{t+1}}$ can be measured in the computational basis to obtain $c$ samples from $\ppi_{t+1}$, allowing to proceed iteratively in the state preparation in the sequence.

By the slowly evolving assumption, the overlap between $\ket{\ppi_t}$ and $\ket{\ppi_{t+1}}$ is constant and amplitude amplification to $\ket{\ppi_{t+1}}$ for $c$ copies in parallel has gate complexity in $\OO\big(c \, \sqrt{\smash[b]{\delta_t'^{-1}}} \, \big)$, where $\delta_t' := \min \{\delta_t,\delta_{t+1}\}$. To simplify the statement of the result we can assume that $\delta_t$ and $\delta_{t+1}$ are multiplicatively close, that is, $\frac{1}{\kappa}\,\delta_t  \leq \delta_{t+1} \leq \kappa \, \delta_t$ for some constant $\kappa > 1$. Hence the time complexity of this final amplitude amplification is in $\OO\big(c\, \sqrt{\smash[b]{\delta_t^{-1}}} \, \big)$, which is dominated by the run-time necessary to initially prepare $\ket{\ppi_{t}}^{\otimes c}$. Consequently, the overall run-time of this process is $\OO\big( c^2 \,C(\ppi) \sqrt{\delta^{-1}} \big)$ per each MC, which is no worse than $\sqrt[4]{N}$, as advertised.

We highlight that the entire procedure only requires classical memory between consecutive time steps in the sequence, in the form of $c$ classical samples stored in memory\footnote{Quantum access to both $P_t$ and $P_{t+1}$ is also assumed, which entails a factor of two increase in the required memory size.}. This is without loss of generality since, as we show later on, $\Omega\big(C(\ppi)\big)$ reflections are needed even if one allows for quantum memory. Moreover, assuming that the classical memory is devoid of errors, this observation that no quantum memory is required shows that approximation errors do not accumulate in the slowly evolving sequence. In fact, the quantum algorithm performed at step $t+1$ does not receive any quantum state as input from step $t$, but only classical information. Of course, each step $t$ still entails a finite failure probability, albeit exponentially small in $c$.

If quantum memory is available, the algorithm can be made slightly more efficient, in terms of how many samples (classical or quantum) are required. Instead of $c$ classical samples, one need store only one coherent sample. The basis of this is a simple near-deterministic cloning algorithm producing two copies of $\ket{\ppi}$ from one, which may be of independent interest. Details of this quantum memory algorithm are provided in \app{quant_mem}.

\section{Application in quantum machine learning}
\label{sec:marked_elem}

We now consider a modification of the $\textsf{Prepare}$ algorithm, which finds application, \textit{e.g.}, in quantization of the reflective Projective Simulation (rPS) model. For the reader interested in quantum ML, details about the rPS can be found in~\cite{2014_Paparo}. Here it is sufficient to point out that the outputs in the rPS model are not samples from $\{\ppi_t\}_t$ but come from restricting to a subset of ``marked elements'' $\M \subseteq \{1,2,\ldots, N\}$ and, typically, the number of marked elements $M = |\M|$ is much smaller than $N$. That is, we want to sample from $\ppim$, the (normalized) probability distribution obtained by restricting $\ppi$ to $\M$:
\begin{align}
	\pi^\M (x) 
	\; := \;
	\begin{cases}
	\frac{1}{\mu} \ \pi (x) & \quad \text{if } x \in \M \\
	0 & \quad \text{if } x \notin \M \;,
	\end{cases}	
\end{align}
where $\mu := \sum_{x \in \M} \pi(x)$ and therefore $\ket{\ppim} = \frac{1}{\sqrt{\mu}} \sum_{x \in \M}\sqrt{\pi(x)} \, \ket{x}$. The set of marked elements is specified by two black-box unitary maps, a \emph{membership oracle} $\mathcal{P}_\M$ and by a \emph{sparse oracle} $\mathcal{Q}_\M$. The former, given an element $x$, specifies whether $x \in \M$ or not; the latter is a quantum accessible memory that, upon input of a number $\nu \in \{1, \ldots, M \}$, outputs the $\nu$-th element of $\M$ (\textit{i.e.}\ $x_\nu \in \M$).

Accessing the oracle $\mathcal{P}_\M$ twice, one can implement a reflection over the subspace of marked elements. This allows to run amplitude amplification, as done, \textit{e.g.},\ in the context of rPS~\cite{2014_Paparo}, to rotate $\ket{\ppi}$ to $\ket{\ppim}$. This operation has a run-time of $\OO \big( \sqrt{\delta^{-1}} \sqrt{\mu^{-1}} \big)$, yielding a quadratic improvement in both mixing time and hitting time with respect to classical methods. This can be done provided that an initial copy of $\ket{\ppi}$ is available. We now explain how, in this context, the state $\ket{\ppi}$ can sometimes be made available more cheaply.

We consider two modified algorithms for preparation of $\ket{\ppi}$: these are equal to the algorithms specified before in \alg{from_unif} and \alg{from_samples}, except for the choice of the initial states, which now are chosen to have support on the marked elements. Specifically, the initial state is either $\ket{\psi_{in}} = \ket{\textbf{u}^\M}$ or $\ket{\psi_{in}} = \ket{x_j}$ for $x_j \in \vec{x}$, where now $\vec{x}$ is a set of $c$ samples drawn previously from $\ppim$. These modified state preparation algorithms require that the new input states $\ket{\psi_{in}}$ can be efficiently produced. This is obviously true for classical samples, while $\ket{\textbf{u}^\M}$ can be prepared with one access to the $\mathcal{Q}_\M$ oracle, since $\ket{\textbf{u}^\M} = \mathcal{Q}_\M \sum_{\nu = 1}^M \frac{1}{\sqrt{M}} \ket{\nu}$. This is then sufficient to perform amplitude amplification of $\ket{\psi_{in}}$ to $\ket{\ppi}$, for both choices of $\ket{\psi_{in}}$. Using similar reasoning as done previously, we see that the amplitude amplification has run-time scaling as
\begin{align}
	\big|\braket{\textbf{u}^\M}{\ppi}\big|^{-1} 
	& \; = \; 
	\sqrt{\mu^{-1}} \ \frac{\sqrt{M}}{ f(\ppim)} \;,
\end{align}
when trying to prepare $\ket{\ppim}$ from $\ket{\textbf{u}^\M}$; and when performing amplitude amplification starting from the available samples $\vec{x}$ the expected run-time is 
\begin{align}
	\mathds{E}_{\ppim}
	\left[\ |\braket{X}{\ppi}|^{-1}\, \right] 
	& \; = \;
	\sqrt{\mu^{-1}} \
	f(\ppim)\;.
\end{align}
Again, combining these two state preparation algorithms (which start from initial states having support on the marked subspace $\M$) into a single procedure we obtain a run-time in $\OO\big( c^2 \, C(\ppim) \sqrt{\mu^{-1}} \, \sqrt{\delta^{-1}} \big)$ for preparing $c$ copies of $\ket{\ppi}$. The preparation from $\M$ is then more efficient whenever $C(\ppim) \sqrt{\mu^{-1}}  < C(\ppi)$; notice that $C(\ppim) \leq \sqrt[4]{M}$, since $\ket{\ppim}$ has support on just $M$ elements.

For the problem of sampling from marked elements, being in a slowly evolving sequence of MCs allows to make the initial set of $c$ samples available. In this case, we again use the state preparation for MC$_t$ to prepare $\ket{\ppi_{t}}^{\otimes c}$ and then map them to $\ket{\ppi_{t+1}}^{\otimes c}$. The final step consists in running the algorithm of~\cite{2014_Paparo} (namely, an amplitude amplification of the marked subspace) in order to obtain $c$ samples from $\ket{\ppim_{t+1}}$. This allows then to proceed inductively with sampling in the sequence. The final projection has $\OO \big(c\, \sqrt{\mu^{-1}} \sqrt{\delta^{-1}}  \big)$ gate complexity and thus is dominated by the cost of preparing $\ket{\ppim_{t+1}}^{\otimes c}$.

We notice, finally, that the method just presented can be directly used to produce new copies of $\ket{\ppim}$, thus directly solving the problem considered in~\cite{2014_Paparo}. It can be straightforwardly realized by running any of the algorithms for preparing $\ket{\ppi}$ followed by amplitude amplification of the subspace of marked elements.

\section{Optimality analysis}
\label{sec:optimality}

To begin with, notice that preparing coherent encodings $\ket{\ppi}$ is in general a difficult task, even when sampling from $\ppi$ can be done efficiently. Consider a randomized algorithm that produces a outcome $x$ with probability $\pi(x)$ which makes a number of binary random choices selecting a computational branch $\text{b}_1$ or $\text{b}_2$ with probabilities $p$ or $1-p$. One can ``purify'' this algorithm to a unitary quantum circuit by substituting every random choice  with a controlled dependence on a pure qubit prepared in the state $\sqrt{p}\,\ket{\text{b}_1} + \sqrt{1-p} \, \ket{\text{b}_2}$. The resulting output state then has the form $\ket{\widetilde{\ppi}} = \ \sum_x \sqrt{\pi(x)}\, \ket{x} \ket{\phi(x)}$ where $\ket{\phi(x)}$ contains residual information of all the choices. Starting from $\ket{\widetilde{\ppi}}$ one cannot directly obtain $\ket{\ppi}$ since there is, in general, no efficient deterministic method that allows one to erase the information contained in the second register. In fact, the possibility to efficiently produce a coherent encoding $\ket{\ppi}$ for all probability distributions which can be efficiently sampled would imply $\textsf{SZK} \subseteq \textsf{BQP}$~\cite{2003_Aharonov}, that is, Statistical Zero Knowledge problems (including, \textit{e.g.},\ Graph Isomorphism) could be solved in quantum polynomial time. While the inclusion $\textsf{SZK} \subseteq \textsf{BQP}$ is not impossible, it is expected that specific structures of the problems have to be exploited (\textit{e.g.},\ graph-theoretic properties in Graph Isomorphism), while the methods based on MC mixing are oblivious to such problem structures. Consequently, it is highly unlikely that any quantum algorithm can prepare $\ket{\ppi}$ encoding stationary distributions of time-reversible MCs in polylog($N$) time, not even when the classical mixing process is fast (\textit{i.e.}, when the MC mixes in polylog($N$) time).

We prove in the \app{lower_bound} that our algorithm is strictly optimal in the class of sampling algorithms which utilize oracle access to reflections about $\ket{\ppi}$ (and do not use other properties of the transition matrix $P$ of the MC) as is the case of many algorithms based on Szegedy quantum walk \cite{2004_Ambainis,2007_Magniez,2010_Krovi,2011_Temme_Nature,2012_Yung}. Specifically, we show that if we start from $c$ copies of $\ket{\ppi}$ and the goal is to obtain $c+1$ classical samples from $\ppi$ (for some constant $c$) then $\Omega( \sqrt[4]{N} )$ accesses to the reflection oracle are required (more tightly, we can prove a $\Omega(\, C(\ppi) \,)$ lower bound). Our proof relies on the ``inner-product adversary'' method developed in the context of quantum money~\cite{2013_Aaronson}, a so-called computational no-cloning theorem.

This $\Omega( \sqrt[4]{N} )$ lower bound actually applies to any MC, also outside of the context of slowly evolving sequences of MCs. Any algorithm that uses quantum walks just to realize the reflection around $\ket{\ppi}$, and then subsequently uses such reflections in a black-box fashion, cannot avoid an $\O(\sqrt[4]{N})$ dependence in its run-time. In particular, algorithms of this type cannot generically achieve the conjectured quadratic speed-up for sampling from stationary distributions of time-reversible MCs~\cite{2007_Richter}. Hence, other techniques are needed.

We finally point out that, however, in our algorithms we have full access to the transition matrix $P$ and, thus, we are not restricted to using reflections around $\ket{\ppi}$. In particular, we can implement a classical random walk as well. If the MC is rapidly mixing then, by definition, the random walk allows to efficiently sample from $\ppi$, while achieving the same goal having access only to reflections around $\ket{\ppi}$ and some initial copies of $\ket{\ppi}$ could take an exponentially longer time.

\section{Discussion}
\label{sec:discussion}

We have presented quantum algorithms for generating samples from stationary distributions of a sequence of Markov chains which achieve a quadratic improvement over previous approaches that can guarantee the generation of the correct output, and work for all time-reversible chains. To achieve this improvement we do not assume special properties of the chain (except detailed balance) but rather we have considered settings where the chains come in a context, namely in a slowly evolving sequence. This result thus has application to all MCs where this framework is natural.

An important domain of application includes statistical physics and material science, where the slowly evolving context, and the need for independent samples, arise when studying phase transitions~\cite{2010_Nishimori}.

A second important family of applications occurs in machine learning (ML), both in the reinforcement learning case~\cite{1998_Sutton} and in the context of generative models~\cite{2016_Bishop}. To briefly comment on this domain, as mentioned earlier in reinforcement learning settings~\cite{1998_Sutton} where the learner's distribution over actions is specified by MCs, the MCs are sequentially updated as the system learns~\cite{2014_Paparo, 2012_Briegel, 2012_Mautner, 2017_Dunjko}. The other facet involves the training of certain generative models (used, \textit{e.g.},\ for unsupervised learning), such as Boltzmann machines~\cite{2012_Fischer}. Here one encounters the need for producing samples from stationary distributions (\textit{e.g.},\ Gibbs states) which are themselves slowly modified as the model is updated~\cite{2008_Tieleman,2014_Wiebe}.

We remark that, in ML, the subsequent Markov chains in the sequence are generated according to a training algorithm which depends on the external outputs of previous Markov chains. Whenever this is the case, the methods developed for quantum-enhanced annealing methods become unsuitable, as they need to keep coherence through the protocol steps~\cite{2008_Wocjan, 2009_Wocjan}.

We conclude observing that, as a feature of our protocol, at each time step we do not output just a classical sample from the target stationary distribution, but a coherent encoding of this distribution. This is not a guaranteed characteristic of quantum mixing protocols~\cite{2007_Richter} and makes our approach suitable for combining with other quantum protocols which start from such a coherent encoding~\cite{2011_Magniez_SIAM, 2014_Paparo, 2010_Krovi, 2015_Montanaro}.

\subsection*{Acknowledgements}
 
The authors acknowledge support by the Austrian Science Fund (FWF) through the SFB FoQuS F4012, the Templeton World Charity Foundation grant TWCF0078/AB46, and the DK-ALM: W1259-N27. 
V.D.\ thanks G.\ D.\ Paparo for initial discussions, and acknowledges the support from the Alexander von Humboldt Foundation. D.O.\ thanks F.\ Guatieri for discussion.
The authors also thank an anonymous reviewer for pointing out a mistake in a previous version of the work, and a method to correct it.

\subsection*{Author contributions} 

V.D.\ and H.J.B.\ wrote a preliminary version of the article. D.O.\ has worked on analysing and extending the algorithm as here presented and has written the current version of the article, all under the supervision of H.J.B.\ and V.D.

\bibliographystyle{unsrt}

\clearpage
\onecolumngrid
\appendix

\section{Markov chain notions}
\label{app:MC}

Here we review the fundamental notions of Markov chain theory and refer to \cite{1998_Norris,2017_Levin} for further details. \\

\paragraph{Transition matrices and probability distributions:} We deal with discrete-time Markov chains having a finite number $N$ of states. Therefore, to a MC is associated a left-stochastic matrix $P$ (a matrix with non-negative entries which add up to one in every column) of size $N\times N$, and each entry $P_{x,y}$ specifies the transition probability from the state $x$ to state $y$. Correspondingly, the non-negative (column) vector $\ppi$ denotes a probability distribution over the state space as
\begin{align}
	& \ppi \; = \;(\, \pi(1), \ldots ,\pi(N) \,)^T 
	\nonumber \\
	& \text{with }~~ \sum_{x=1}^N \pi(x) \; = \; 1 \;.
\end{align}
A MC is then specified by a the transition matrix $P$ and an initial probability distribution $\ppi_{in}$. We stick to the convention of left-stochastic matrices which act from the left on column vectors $\ppi$ representing probability distributions, that is $\ppi' = P \ppi$. This convention is not customary in the MC literature (where the usage of right-stochastic matrices prevails), but it matches the one adopted in the quantum information community. In particular, $P_{y,x}$ denotes the transition probability from the element $x$ to the element $y$.\\

\paragraph{Ergodic MCs:} A $N$-state MC is \emph{irreducible} if it is possible from each state $x$ to reach any other state $y$ in a finite number of steps and with non-zero probability. The \emph{period} of a state $x$ is the largest positive integer such that any return to $x$ can occur only at multiples of that integer. If the period of all states is 1, the MC is said to be \emph{aperiodic}. If $P$ is irreducible and aperiodic, then there exists a unique \emph{stationary distribution} $\ppi$, such that:
\begin{align}
	P \ppi \; = \; \ppi \;
\end{align}
and, moreover, $\ppi$ has support over all the elements of the MC. This also implies that, under application of a sufficiently large number of steps any initial probability distribution $\widetilde{\ppi}$ will converge to the unique stationary distribution, $\lim_{k\rightarrow\infty} P^k \widetilde{\ppi} = \ppi$. This convergence process is called \emph{mixing}, and since MCs mix if and only if they are irreducible and aperiodic, these are called \emph{ergodic Markov chains}. \\

\paragraph{Time reversal:} The \emph{time reversal} $\widehat{P}$ of a Markov chain $P$ having stationary distribution $\ppi$ is defined as:
\begin{align}
	\widehat{P}_{y,x} \, := \, P_{x,y} \ \frac{\pi(y)}{\pi(x)}
\end{align}
and a MC is said to be time-reversible if it is equal to its time-reversed version, $P = \widehat{P}$. Equivalently, a time-reversible MC is one that satisfies the \emph{detailed balance equation}:
\begin{align}
	P_{y,x} \, \pi(x) \; = \; P_{x,y} \, \pi(y) \;.
\end{align}
We can also write the time reversed MC in matrix form as $\widehat{P} = D(\ppi) P^T D(\ppi)^{-1},$ where $D(\ppi)$ is the diagonal matrix $D(\ppi) := \text{diag}(\, \pi(1), \ldots, \pi(N) \,)$. This implies that if $P$ is time reversible, then its spectrum is real. In the following, we will always consider ergodic and time-reversible MCs. \\

\paragraph{Mixing times:} Obviously, not all mixing process of ergodic MCs are equally fast. We use the \emph{total variation distance}, defined as $d(\ppi,\ppi') := \frac{1}{2}\sum_x |\pi(x) - \pi'(x)|$ to assess the speed of mixing (the total variation distance exactly matches the trace distance in the quantum information context). We then define $d(k):= \max_{\boldsymbol{\sigma}} d(P^k \boldsymbol{\sigma}, \ppi)$ as the distance in distributions between a sample drawn after $k$ walk steps starting from any distribution $\boldsymbol{\sigma}$ and stationary distribution $\ppi$ of $P$. The \emph{mixing time} $t_\text{mix}(\epsilon)$ then is defined as the smallest time necessary to bring any initial distribution within distance $\epsilon$ from the stationary distribution, $d(t_\text{mix}(\epsilon)) \leq \epsilon$. We then set $t_\text{mix} := t_\text{mix}(1/4)$. It can be shown then that the convergence of an ergodic MC is exponentially fast in terms of the mixing time, that is:
\begin{align}
	d(\ell t_\text{mix}) 
	\; \leq \; 
	2^{-\ell} \;.
\end{align}
The mixing times often play the critical role in the computational complexity of MC-based algorithms. There are many techniques that can be employed for upper and lower bounding the mixing time, but one of the most useful characterizations is the following. Because of Perron-Frobenius theorem all eigenvalues of a left-stochastic matrix $P$ are smaller or equal to 1 in modulus. If $P$ is ergodic then its stationary distribution $\ket{\ppi}$ is the only eigenvector of $P$ having eigenvalue equal to $+1$. That is, all other eigenvectors have eigenvalues $\lambda$ with $|\lambda| < 1$. Let $\sigma(P)$ be the spectrum of a time-reversible Markov chain $P$; we define the \emph{spectral gap} $\delta$ of $P$ as:
\begin{align}
		\delta 
		\; := \; 
		1 - \max_{\lambda\in \sigma(P): \atop \lambda \neq 1} | \lambda |
\end{align}
\textit{i.e.}\ the minimum of $1 - |\lambda|$ over the eigenvalues of $P$ which differ from one. The spectral gap is a rather tight estimate for the inverse of the mixing time, since
\begin{align}
	\left(\frac{1}{\delta} -1 \right) \log \left(\frac{1}{2\epsilon} \right)
	\ \leq \
	t_\text{mix}  (\epsilon)
	\ \leq \	
	\frac{1}{\delta} \log\left( \frac{1}{\epsilon \, \pi_\text{min}} \right)
\end{align}
holds for all time-reversible MCs (where $\pi_\text{min}$ is the smallest probability in $\ppi$). In short we have $t_\text{mix} \in \OO(1/\delta)$ and 
$1/\delta \in \OO(t_\text{mix})$, giving asymptotic upper and lower bounds to the mixing time.

\section{Szegedy quantum walk}
\label{app:Szegedy}

Here we review the basics of Szegedy quantum walks~\cite{2004_Szegedy}. For further details see~\cite{2010_Krovi, 2011_Magniez_SIAM} and references therein. \\

\paragraph{Szegedy walk operator:} The Szegedy walk operator $W(P)$ can be implemented for any transition matrix $P$, and not only for those associated to ergodic and time-reversible MCs (but $W(P)$ has nice spectral properties only if $P$ is ergodic and time reversible). The basic building block to define $W(P)$ is the \emph{diffusion operator} $U_P$ which acts on two quantum registers of $N$ states and is (partially) defined as follows:
\begin{align}
\label{eq:UP}
	U_{P} \ket{x}_1\ket{0}_2 
	& \; := \; 
	\ket{x}_1 \sum_{y=1}^{N}\sqrt{P_{x,y}} \, \ket{y}_2 \;.
\end{align}
By measuring the second register in the computational basis a step of the classical random walk is obtained, hence $U_P$ is a natural way of defining a quantum extension of the classical MC. When we say that we have quantum access to $P$, we specifically mean that we have access to a diffusion operator of the form~\eqref{eq:UP}. The diffusion operator $U_P$ can be efficiently realized, for instance, when $P$ is a sparse transition matrix. Then, we can define the \emph{Szegedy walk operator} as the unitary
\begin{align}
	W(P) \; := \;   \textsc{Swap} \; U_P \, (\II_1 \otimes Z_2) \; U_P^\dag \;,
\end{align}
where $Z_2 := 2 \ketbra{0}_2 - \II$ and $\textsc{Swap}$ interchanges the first and second register. $W(P)$ acts non-trivially on the invariant subspace $A+B$, where $A := \textup{span} \{\, U_P \ket{x}\ket{0} \,\}_x$ and $B := \textup{span} \{\, \textsc{Swap} \; U_P \ket{x}\ket{0} \,\}_x$. \\

\paragraph{Spectral properties of $W(P)$:} When $P$ is ergodic and time-reversible the space $A+B$ has dimension $2N-1$ and the intersection $A \cap B$ contains only the state $U_P \, \ket{\ppi,0} = \textsc{Swap} \; U_P \ket{\ppi,0}$, as one can verify using the detailed balance equation for $P$. The state $U_P \, \ket{\ppi,0}$ is the only $+1$-eigenstate of $W(P) \Pi_{A+B}$, where $\Pi_{A+B}$ is a projector on the invariant subspace $A+B$. Moreover on the invariant subspace the other $2N-2$ eigenvalues of $W(P)$ are given by $\{ e^{\pm i \theta_\ell} \}_{\ell \in [N]}$ where $\{ \cos \theta_\ell \}_{\ell \in [N]}$ are the eigenvalues of $P$ that are different from one (remember, the spectrum of $P$ is real for time-reversible MCs). With this notation the phase gap ($\Delta$) and spectral gap ($\delta$) are given by 
\begin{align}
	\begin{cases}
	   \Delta \; := \; \underset{\ell}{\min} \ \{\, \theta_\ell \,\}  \\
	\; \delta \; := \; \underset{\ell}{\min} \ \{\, 1 - |\cos \theta_\ell| \,\}
	\end{cases}	
	\quad \text{with}~ \theta_\ell \in (0 , \pi)	
\end{align}
and therefore the phase gap is quadratically larger than the spectral gap, $\Delta \geq \sqrt{2 \delta}$.

\section{Subroutines based on quantum walks}
\label{app:subroutines}

Here we show how to use Szegedy walk operator within the phase detection algorithm to implement projective measurements onto $\ket{\ppi}$ and partial reflections around $\ket{\ppi}$, as originally done in~\cite{2011_Magniez_SIAM}; we also show how to use fixed-point amplitude amplification to deterministically map a given input state to $\ket{\ppi}$. \\

\paragraph{Phase estimation and phase detection:} The phase detection algorithm applied to the Szegedy walk operator $W(P)$ is illustrated in \fig{2} and its action is as follows. Call $\{ \ket{\theta_\ell} \}_\ell$ the eigenvectors of $W(P)$ having eigenvalue $e^{i\theta_\ell}$, $W(P) \ket{\theta_\ell} = e^{i\theta_\ell} \ket{\theta_\ell}$. In particular $\ket{\theta_\ell = 0} \equiv \ket{\ppi}$. Then, given an input state of the form $\ket{\psi_{in}} = \sum_{\ell} \psi_{\theta_\ell} \, \ket{\theta_\ell}$ the phase detection algorithm outputs an approximation of the state
\begin{align}  \label{eq:PE}
	\ket{\psi_{out}} 
	\; = \;
	\psi_0 \, \ket{\ppi} \ket{1}+
	\sum_{\ell:\; \theta_\ell \neq 0} 
	\psi_{\theta_\ell} \, \ket{\theta_\ell} \ket{0} \;.
\end{align}
That is, the second register contains a bit signalling whether the $\ket{\theta_\ell}$ is the $+1$-eigenvector or not. The phase detection algorithm produces a state within trace distance $\err$ from the state in Eq.~\eqref{eq:PE} using $\O \left( \Delta^{-1} \log \err^{-1} \right) = \OO \big( \sqrt{\delta^{-1}} \big)$ oracle accesses to controlled-$W(P)$ and $\O \left( \Delta^{-1} \log \err^{-1} \right) = \OO \big( \sqrt{\delta^{-1}} \big)$ extra gates. In particular, the cost has only a logarithmic dependence on the error, see \textit{e.g.},~\cite{2011_Magniez_SIAM, 2013_Svore,2016_Wiebe}. \\

\begin{figure*}
\includegraphics[width=\textwidth]{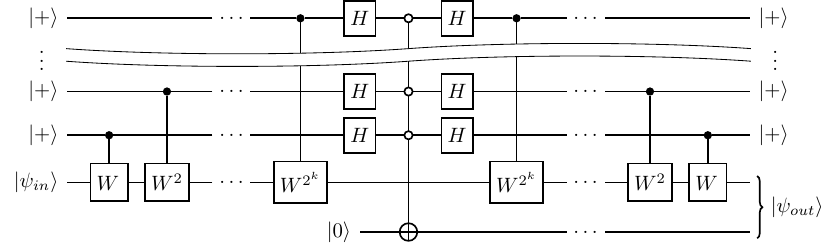}
\caption{Phase detection algorithm applied to the Szegedy operator $W(P)$. The left part of the circuit implements the standard phase estimation algorithm, except for the fact that a final Hadamard transform is applied instead of an inverse quantum Fourier transform. Using the Hadamard transform is sufficient since we only require to discriminate the $+1$ eigenvector of $W(P)$ from eigenvectors having eigenvalue different from $+1$. The central multi-controlled Toffoli gate flips an ancilla qubit if and only if all control qubits are in $\ket{0}$. The right part of the circuit finally uncomputes the value contained in the ancillary registers.}
\label{Fig2}
\end{figure*}

\paragraph{Projective measurement onto $\ket{\ppi}$:} The phase detection algorithm allows to directly implement the projective measurement given by the projectors $\big\lbrace \ketbra{\ppi}, \, \II - \ketbra{\ppi} \big\rbrace$, applied to any arbitrary input state $\ket{\psi_{in}}$: it is realized by measuring the second register of the state in Eq.~\eqref{eq:PE} in the computational basis. The gate and oracle complexity of this projective measurement is thus the same of the phase detection algorithm, $\OO \big( \sqrt{\delta^{-1}} \big)$. The success probability of the measurement, applied on an input pure state $\ket{\psi_{in}}$, is approximately $|\braket{\psi_{in}}{\ppi}|^2$. Moreover, the classical outcome of this projective measurement is a bit that signals whether the projection onto $\ketbra{\ppi}$ was successful or not. This allows, \textit{e.g.},\ to redo the preparation and measurement process until the algorithm succeeds in obtaining the target state $\ket{\ppi}$. \\

\paragraph{Partial reflections around $\ket{\ppi}$:} Next, the phase detection algorithm can be used to approximately implement the \emph{partial reflection}
\begin{align}
	\textup{R}_\phi(\ppi) 
	\; := \; e^{i\phi} \ketbra{\ppi} \;+\; \big(\, \II - \ketbra{\ppi} \,\big) \;
\end{align}
where $\phi$ is a tunable parameter. Notice that for $\phi = 180^\circ$ the partial reflection becomes a standard reflection around $\ket{\ppi}$, $\textup{R}(\ppi) = \II - 2\ketbra{\ppi}$. A partial reflection can be implemented using once the circuit in \fig{2} and once its inverse: with the first call a input state $\ket{\psi_{in}}$ is mapped to a state as in Eq.~\eqref{eq:PE}; then, a phase $e^{i\phi}$ is applied selectively on the ancilla qubit being in the $\ket{1}$ state; finally, the phase detection algorithm is run in reverse to uncompute the bit contained in the second register. In summary:
\begin{align} \label{partial_reflection}
	\ket{\psi}\ket{0}
	& \ \mapsto \
	\psi_0 \, \ket{\ppi} \ket{1}+
	\sum_{\ell:\; \theta_\ell \neq 0} 
	\psi_{\theta} \, \ket{\theta_\ell} \ket{0} 
	\nonumber\\
	& \ \mapsto \
	e^{i\phi} \ \psi_0 \, \ket{\ppi} \ket{1}+
	\sum_{\ell:\; \theta_\ell \neq 0} 
	\psi_{\theta} \, \ket{\theta_\ell} \ket{0}
	\nonumber\\
	& \ \mapsto \
	\bigg(
	e^{i\phi} \ \psi_0 \, \ket{\ppi}+
	\sum_{\ell:\; \theta_\ell \neq 0} 
	\psi_{\theta} \, \ket{\theta_\ell} 	
	\bigg) \ \ket{0} \;.
\end{align}
Notice that in the operation given above we apply in sequence a phase estimation and its inverse, and these two operations cancel out. In conclusion, the oracle and gate cost of approximating a partial reflection is also in $\O \big( \Delta^{-1} \log \err^{-1} \big) = \OO \big( \sqrt{\delta^{-1}} \big)$. \\

\paragraph{Fixed-point amplitude amplification:} Partial reflections are fundamental for implementing \emph{fixed-point amplitude amplification} ($\FPAA$). This is a variant of amplitude amplification whereby the output state can get arbitrarily close to the ideal target state. In standard amplitude amplification usually one has the ``souffl\'e problem''~\cite{2005_Grover}: if the rotation in the amplitude amplification process is not stopped at the right moment, the fidelity with the target state starts decreasing again; moreover, even using the optimal number of reflections, only a constant fidelity between the outputs state and the ideal target is reached. In contrast, in $\FPAA$ one gets exponentially close to the target state $\ket{\psi_{out}}$, allowing an almost exact preparation of this state, with only a logarithmic dependence of run-time on the approximation error. Details follow.

A $\FPAA$ algorithm takes a single copy of a input state $\ket{\psi_{in}}$ and maps it to a state $\err$-close to $\ket{\psi_{out}} \equiv \Pi_{out} \ket{\psi_{in}} / \norm{\Pi_{out} \ket{\psi_{in}}}$, where $\Pi_{out}$ is a projector over a target subspace (after the process the input state $\ket{\psi_{in}}$ is no longer available). More precisely, in order to implement $\FPAA$ three ingredients are required: 
\begin{enumerate}[topsep=4pt,itemsep=0pt,partopsep=4pt,parsep=2pt]
\item a single copy of a input quantum state $\ket{\psi_{in}}$;
\item the ability of implementing \emph{partial reflections} around the input state;
\item the ability of implementing \emph{partial reflections} around the target subspace.
\end{enumerate}
Specifically, these partial reflections are respectively given by $\textup{R}_\phi(\psi_{in}) = e^{i\phi} \ketbra{\psi_{in}} + \big( \II - \ketbra{\psi_{in}} \big)$ and $\textup{R}_{\phi'}(\Pi_{out}) = e^{i\phi'} \ \Pi_{out} + \big( \II -  \, \Pi_{out} \big)$, for arbitrary angles $\phi,\phi'$.

The $\FPAA$ algorithm of Yoder et.~al.~\cite{2014_Yoder} can be implemented, provided that the conditions 1.-3.\ hold, and has both the quadratic speedup of Grover search and the fixed-point property. This algorithm is parametric, depending on two input parameters, $\err \in [0,1]$ and $\gamma \in (0,1)$: if $|\braket{\psi_{out}}{\psi_{in}}| \geq \sqrt{\gamma}$ holds, then the output of the $\FPAA$ algorithm is a state with $\err$ distance in trace norm from the ideal $\ket{\psi_{out}}$. The number of calls to $\textup{R}_\phi(\psi)$ and $\textup{R}_{\phi'}(\Pi_{out})$ is in $\O(\sqrt{\gamma^{-1}} \log \err^{-1})$. \\

\paragraph{Heralded state preparation:} We finally show how to use $\FPAA$ followed by a projective measurement to implement (in an efficient way) a \emph{heralded preparation} of $\ket{\ppi}$. If we start from an initial state $\ket{\psi_{in}}$ and then apply the projective measurement $\big\lbrace \ketbra{\ppi}, \, \II - \ketbra{\ppi} \big\rbrace$ the success probability is $|\braket{\ppi}{\psi_{in}}|^2$; but the success probability of the measurement process can be increased by preceding the measurement by a round of amplitude amplification. On average, the procedure using amplitude amplification has a quadratically smaller run-time in producing a copy of $\ket{\ppi}$.

More precisely, the heralded state preparation works as follows. We first run the optimal $\FPAA$ algorithm~\cite{2014_Yoder} using reflections around a initial state $\ket{\psi_{in}}$ and around the target state $\ket{\ppi}$; the algorithm is run setting $\err$ as the target approximation error and setting some value $\gamma > 0$ as overlap parameter. This means that, if the inequality $|\braket{\ppi}{\psi_{in}}|^2 \geq \gamma$ holds, then $\FPAA$ guarantees to output a state within $\err$ distance from $\ket{\ppi}$; however, if instead $|\braket{\ppi}{\psi_{in}}|^2 < \gamma$ holds, the output state can be arbitrarily far from $\ket{\ppi}$. To ameliorate this issue, after the $\FPAA$ we apply a projective measurement onto $\ket{\ppi}$ (or onto the orthogonal subspace). When the measurements succeeds, the preparation of (a approximation of) $\ket{\ppi}$ is guaranteed, independently from the initial overlap $|\braket{\ppi}{\psi_{in}}|$. We also remind that this final measurement succeeds almost deterministically (with probability $1-\err$) when $|\braket{\ppi}{\psi_{in}}|^2 \geq \gamma$ holds. The number of reflections needed in the heralded preparation of $\ket{\ppi}$ is then $\O(\sqrt{\gamma^{-1}} \log (\err^{-1}))$ and the total run-time is in $\OO(\sqrt{\gamma^{-1}} \log^2 (\err^{-1}) )$, as we will prove in the next Appendix.

\section{Analysis of imperfect reflection operators}
\label{app:imperfect_ref}

Here we consider the propagation of errors when the partial reflection used within $\FPAA$ are approximate and how the run-time is affected. For this section only, we assume that the relevant parameters in the soft-$\O$ notation are $\delta, \gamma$ and $\log \err^{-1}$; namely, we keep $\log \err^{-1}$ terms and discard $\log \log \err^{-1}$ dependencies.\\

We suppose that $\sqrt{\gamma}$ is (a lower bound to) the overlap between $\ket{\psi_{in}}$ and $\ket{\ppi}$, and the final targeted error is $\err$. $\FPAA$ entails the use of $\O \big( \sqrt{\gamma^{-1}} \log \err^{-1} \big)$ perfect reflections in order to achieve the desired accuracy goal. However, the same can be achieved with imperfect reflections, provided that each reflection has an error smaller of $\err /\text{(number of steps)}$: by the triangle inequality the total accumulated error will be upper bounded by $\err$. That is, we need to implement a partial reflection with an accuracy
\begin{align}
	\err^\textup{R}
	\; = \; 
	\O \left(
	\frac{ \err }{ \sqrt{\gamma^{-1}} \log \err^{-1}}
	\right) \;.
\end{align}
Hence the total gate cost of the heralded state preparation procedure (nesting approximate reflections within $\FPAA$) is given by:
\begin{align}
	& \O \left( \sqrt{\gamma^{-1}} \log \err^{-1} \right) 
	\times
	\O \left( \sqrt{\delta^{-1}} \ \log\big(1/ \err^\textup{R}\big) \right)
	\nonumber\\
	\ = \ 
	& \O \left( \sqrt{\gamma^{-1}} \, \sqrt{\delta^{-1}}
	\log \left(\err^{-1}\right)
	\left[
	\log \sqrt{\gamma^{-1}} + \log \err^{-1} + \log \log \err^{-1} 
	\right]
	\right) 
	\nonumber\\
	\ = \ 
	& \OO \left( 
	\sqrt{\gamma^{-1}} \, \sqrt{\delta^{-1}} 
	\log^2 \left(\err^{-1} \right) 
	\right) \;.
\end{align}
Thus, heralded preparation of $\ket{\ppi}$ within $\err$ final precision has a overall gate complexity scaling as $\log^2(\err^{-1})$. We also remark that errors do not propagate from one time step to the next in the slowly evolving sequence, since at each time step we freshly prepare new copies of $\ket{\ppi}$. This is possible since we have access to projectors onto the required states, which allow to decrease approximation errors.

\section{Failure probabilities of preparation from uniform and from samples}
\label{app:fail_prob}

We here show that the $\textsf{Prepare}$ subroutines are not overly sensitive to small imperfections and that failure probabilities decrease exponentially with $c$, the number of classical samples carried over in each step of the slowly evolving sequence. \\

\paragraph{Preparation from uniform distribution:} For $\PFU$, as given in \alg{from_unif}, the analysis is simple, since the input state $\ket{\textbf{u}}$ has no error. The algorithm succeeds when at least $c$ out of $2c$ heralded preparations of $\ket{\ppi}$ starting from $\ket{\textbf{u}}$ are successful. In the case in which $\chi' \geq \sqrt{N}/f(\ppi)$ the global probability of failure is $2^{-\O(c)}$. In fact the $2c$ runs have independent outcomes; then, we can apply the Chernoff bound:
\begin{align}
	\Pr[\ \textup{number of failures} \geq (1+\delta)\, 2 \err c \ ]
	\ \leq \
	\exp \left(- \,\frac{\delta^2}{2+\delta} \ 2 \err c \right)
\end{align}
where $\err$ is an upper bound to the failure probability in the preparation of $\ket{\ppi}$ and $\delta > 0$ is a free parameter. Choosing $1 + \delta = 1/(2\err)$ we get:
\begin{align}
	\Pr[\ \textup{number of failures} \geq c \ ]
	\ & \leq \
	\exp \left(- \,\frac{(1-2\err)^2}{1+2\err} \ c \right)
	\nonumber\\
	\ & \leq \
	\exp \left(- \, 0.9 \ c \right) \;,
\end{align}
where the second inequality holds for sufficiently small $\err$. \\

\paragraph{Preparation from samples:} A similar analysis holds for $\PFS$, as given in \alg{from_samples}. Notice that the input samples $\vec{x} = \{x_1, \ldots,x_c\}$ are not (exactly) distributed with $\ppi$, but with a distribution $\tilde{\ppi}$ which is $\err$-close to $\ppi$, say, in total variation distance. Considering a random variable $\tilde{X}$ distributed as $\tilde{\ppi}$ we have, for any $v>0$:
\begin{align}
	\Pr\left[\ \pi^{-1/2}\big(\tilde{X}\big) \geq v \ \right]
	\ & = \
	\sum_{x:~\pi^{-1/2}(x) \,\geq\, v} \tilde{\pi}(x)
	\nonumber\\
	& \leq \
	\err + \sum_{x:~\pi^{-1/2}(x) \,\geq\, v} \pi(x)
	\nonumber\\
	& = \
	\err + \ \Pr\left[\ \pi^{-1/2}\big( X \big) \geq v \ \right]
	\nonumber\\
	& \leq \
	\err + \ \frac{ \mathbb{E}\big[\, \pi^{-1/2}(X)\,\big] }{v} \;.
\end{align}
In the first inequality we have applied the definition of total variation distance and in the second Markov's inequality. Thus, we have 
\begin{align}
	\Pr\left[\ \pi^{-1/2}\big(\tilde{X}\big) 
	\geq 
	2 \,\mathbb{E}\big[\, \pi^{-1/2}(X)\,\big] \ \right]
	\ & \leq \ 
	\err + \frac{1}{2} \;.
\end{align}
Thus, with high probability at least one sample in $x_\ast \in \vec{x}$ satisfies $\pi^{-1/2}(x_\ast) < 2 \, \mathbb{E}\big[\, \pi^{-1/2}(X)\,\big] = 2 f(\ppi)$; namely, this happens with probability at least:
\begin{align}
	1 - \left( \frac{1 + 2\err}{2}\right)^c = 1 - 2^{-\O(c)} \;.
\end{align}
Then, we consider a heralded state preparation of $\ket{\ppi}$ starting from $\ket{x_\ast}$ for $2c$ times. If $\chi \geq f(\ppi)$ the analysis proceeds exactly as the one performed for $\PFU$ and, hence, with probability $1 - 2^{-\O(c)}$ at least $c$ of these $2c$ runs will be successful in producing approximations of $\ket{\ppi}$. The global failure probability of $\PFS$ is thus in $2^{-\O(c)}$. \\

\paragraph{Combined algorithm:} The algorithms $\PFU$ and $\PFS$ are used as subroutines of the combined state preparation algorithm, as specified in the main text. In this algorithm the values of $\chi$ and $\chi' = \sqrt{N}/\chi$ are doubled until they exceed $2 \sqrt[4]{N}$, in which case either $\chi \geq f(\ppi)$ or $\chi' \geq \sqrt[4]{N} / f(\ppi)$ is satisfied: then, the algorithm has to succeed, except with probability $2^{-\O(c)}$.

\section{Quantum memory algorithm}
\label{app:quant_mem}

Here we show that, if a long-term quantum memory is available, only one quantum sample (\textit{i.e.}, $\ket{\ppi_t}$) has to be stored in memory between consecutive steps in the slowly evolving sequence. That is, we assume that the quantum state $\ket{\ppi_t}$ does not decohere during the time in which $P_t$ is updated to $P_{t+1}$. Then, one can store a single copy $\ket{\ppi_t}$ and employ it to (almost deterministically) prepare two copies of $\ket{\ppi_t}$. One copy of $\ket{\ppi_t}$ is provided as external output, while the other copy is rotated to $\ket{\ppi_{t+1}}$ using fixed-point amplitude amplification and $\ket{\ppi_{t+1}}$ is provided as input to the successive MC. Therefore, we only have to show how to implement this state duplication algorithm. \\

\paragraph{State duplication algorithm:} The state duplication algorithm works as follows, assuming that $f(\ppi)$ is known. If $f(\ppi) \geq \sqrt[4]{N}$, then we use $\PFU$ and the second copy of $\ket{\ppi}$ is produced \emph{de novo} from the uniform distribution. Else ($f(\ppi) < \sqrt[4]{N}$), we employ $U_\textsf{PFS}$, a coherent version of $\PFS$ as described in \alg{from_samples}, for the case $c=1$. Notice that $U_\textsf{PFS}$ is a quantum algorithm that tries to prepare $\ket{\ppi}$ from a single classical sample $x$ drawn from $\ppi$; hence it can be written as an isometry (that is, as a unitary operation, plus the ability to add ancillary quantum systems) acting on a register initialized in $\ket{x}$: 
\begin{align}
	U_\textsf{PFS} \ket{x} 
	\; = \;
	\ket{x} 
	\Big[ \sqrt{p_{succ}(x)} \, \ket{\ppi} \ket{\textup{ok}} +
	\sqrt{1- p_{succ}(x)} \, \ket{\psi_x} \ket{\textup{err}} \Big]
	\;.
\end{align}
Here the first register is the control (input) register, the third register contains a flag heralding the successful preparation of $\ket{\ppi}$, and the second register either contains $\ket{\ppi}$ or an arbitrary state $\ket{\psi_x}$ in case of failure. The algorithm $U_\textsf{PFS}$ has a run-time proportional to $f(\ppi)$ and has, averaging on $x$, a constant success probability, say larger than $1/2$: $\mathds{E}_{\ppi} [\, p_{succ}(X)\, ] = \sum_x \pi(x) \, p_{succ}(x) \geq \frac{1}{2}$. We then consider the following quantum computation
\begin{align}
	\ket{\ppi}
	\ & \mapsto \
	\sum_x \sqrt{\pi(x)} \, \ket{x}\ket{x}
	\nonumber\\
	\ & \mapsto \
	\sum_x \sqrt{\pi(x)} \; U_\textsf{PFS}\ket{x} \, U_\textsf{PFS}\ket{x} \;.
	\label{eq:Ux-Ux}
\end{align}
The state in Eq.~\eqref{eq:Ux-Ux} can be rewritten as follows, after rearrangement of the quantum registers:
\begin{align}
	\vert\,\widetilde{\ppi}^{(2)}\,\rangle
	\ := \
	\left[
	\sum_x \sqrt{\pi(x)} \; p_{succ}(x) \, \ket{x}^{\otimes 2}
	\right]	
	\, \ket{\ppi}^{\!\otimes 2} \ket{\textup{ok}'} \, + \,
	\sqrt{1 - p_{succ}'} \, \ket{\psi} \ket{\textup{err}'} \;.
\end{align}
Here the rightmost register is in $\ket{\textup{ok}'}$ if both instances of $U_\textsf{PFS}$ have raised a success flag and is in $\ket{\textup{err}'}$ otherwise; also, the probability of two successful preparations of $\ket{\ppi}$ is given by 
\begin{align}
	p_{succ}' 
	\ = \
	\Big\lVert \sum_x \sqrt{\pi(x)} \; p_{succ}(x) \, \ket{x}^{\otimes 2} \Big\rVert^2
	\ = \
	\mathds{E}_{\ppi} [\, p_{succ}^{\, 2}(X)\, ] 
	\ \geq \ 
	\frac{1}{4} \;.
\end{align}
Hence upon measurement of the last register of $\ket{\tilde{\ppi}^{(2)}}$, when we obtain as outcome $\ket{\textup{ok}'} \equiv \ket{\textup{ok}}\ket{\textup{ok}}$ we also obtain two copies of $\ket{\ppi}$. This happens with probability larger than $1/4$. The final step of the algorithm is to use $\FPAA$ to deterministically apply the projector $\Pi_{\textup{ok}} := \one \otimes \ketbra{\textup{ok}'}$ to the state $\ket{\tilde{\ppi}^{(2)}}$, thus deterministically recovering $\ket{\ppi}^{\otimes 2}$ (together with an ancillary register in a separable quantum state, which can be discarded). Here $\FPAA$ can be implemented using $\OO(1)$ accesses to $U_\textsf{PFS}$ and thus has essentially the same run-time as the classical-memory $\PFS$ algorithm. \\

\paragraph{Further remarks:} Notice that if we want to output $c \geq 3$ copies of $\ket{\ppi}$, this can be obtained by applying the state duplication algorithm in sequence many times (which is more efficient than using a modification of Eq.~\eqref{eq:Ux-Ux} in which $U_\textsf{PFS}$ is used $c$ times in parallel). Finally, if the value of $f(\ppi)$ is not known, one can simply revert to the classical-memory strategy, at the cost of carrying over $c$ classical samples in order to have a $2^{-\O(c)}$ failure probability.

\section{Lower bound on the oracle cost of sampling}
\label{app:lower_bound}

Here we prove a lower bound on the number of reflections around $\ket{\ppi}$ that are needed to produce two classical samples drawn from $\ppi$, starting from a single copy of $\ket{\ppi}$. The lower bound also applies when $c+1$ classical samples have to be produced from $c$ copies of $\ket{\ppi}$, thus we directly prove this more general case. \\

The state preparation algorithms presented in this work allow to prepare $\ket{\ppi}$ using $\OO\big(C(\ppi)\big) \leq \O(\,\sqrt[4]{N}\,)$ reflections around $\ket{\ppi}$, using at most logarithmically many copies of $\ket{\ppi}$. A result of Aaronson and Christiano~\cite{2013_Aaronson} is that there exists a class of states with all positive real amplitudes (effectively, coherent encodings of probability distributions) which require, on average, $\Omega(\sqrt[4]{N})$ accesses to the reflection oracle in order to be duplicated. This already shows that our algorithms have essentially optimal worst-case performance.

We strengthen the result in two ways. First, we prove a $\Omega\big(C(\ppi)\big)$ lower bound in the number of oracle accesses needed, thus matching the $\O\big(C(\ppi)\big)$ oracle complexity attained by our algorithms, for all values of $C(\ppi)$. Secondly, we show that the same lower bound holds also for classical sampling problems. Namely, suppose that we have $c$ initial copies of $\ket{\ppi}$ and the ability to implement controlled reflections around $\ket{\ppi}$, while the goal is to obtain $c+1$ classical samples distributed according to $\ppi$. We show that in order to accomplish this task $\Omega\big(C(\ppi)/ c^2\big)$ controlled-reflection around $\ket{\ppi}$ are required. This means that our algorithm is asymptotically optimal (up to polylogarithmic factors) in the number of queries to a reflection oracle.

The proof of this lower bound hinges upon the ``inner-product adversary'' method~\cite{2013_Aaronson}. We can condense the results of Section 4.2 and Appendix B of~\cite{2013_Aaronson} into the following theorem.

\begin{theo}
\label{thm:Aaronson}
Suppose that we have access to reflection oracles $U_\psi$ (and to its controlled version c-$U_\phi$) so that:
\begin{align}
	U_\psi \ket{\psi} & = - \, \ket{\psi} \nonumber\\ 
	U_\psi \ket{\eta} & = + \, \ket{\eta}	\qquad \forall \, \ket{\eta} \text{ orthogonal to } \ket{\psi} \;.
\end{align}
The states $\ket{\psi}$ come from a subset $\mathcal{Z}$ of the entire Hilbert space $\hil$. Moreover we require that on these states there is a symmetric binary relation $\mathcal{R} \subseteq \mathcal{Z} \times \mathcal{Z}$ such that 
\begin{align}
	\forall \ket{\psi} \in \mathcal{Z}: \quad 
	& (\psi,\psi)  \notin \mathcal{R} \\
	\forall \ket{\psi} \in \mathcal{Z}, \ 
	\exists \ket{\phi} \in \mathcal{Z}: \quad 
	& (\psi,\phi) \in \mathcal{R} \;.	
\end{align}
Suppose, next, that for all $\ket{\psi} \in \mathcal{Z}$ and for all $\ket{\eta} \in \mathcal{H}$ that are orthogonal to $\ket{\psi}$ the following inequality holds
\begin{align}
\label{eq:gamma}
	\underset{\phi \in \mathcal{Z}: \atop (\psi,\phi) \in \mathcal{R}}
	{\mathds{E}}
	\Big[ \ \big\vert
	\braket{\eta}{\phi}
	\big\vert^2 \; \Big]
	\; \leq \; 
	\gamma
\end{align}
for some $\gamma \in \mathds{R}^+$. 

Then, consider a quantum circuit $Q^\psi$ consisting of a fixed set of unitary operations $Q^\ast$ that make oracle calls to c-$U_\psi$ (and similarly, $Q^\phi$ is obtained when $Q^\ast$ calls c-$U_\phi$). Suppose that for all $(\psi,\phi) \in \mathcal{R}$ the quantum states $\big\vert\, \Psi_{in}^\psi \,\big\rangle$ and $\big\vert\, \Psi_{in}^\phi \,\big\rangle$ are two input states such that $\big\vert \big\langle \, \Psi_{in}^\phi \, \big\vert \, \Psi_{in}^\psi \big\rangle \big\vert \geq \alpha$, while the output states $\big\vert \, \Psi_{out}^\psi \, \big\rangle = Q^\psi \big\vert \, \Psi_{in}^\psi \, \big\rangle$ and $\big\vert \, \Psi_{out}^\phi \, \big\rangle = Q^\phi \big\vert \, \Psi_{in}^\phi \, \big\rangle$ have to satisfy $\big\vert \big\langle \, \Psi_{out}^\phi \, \big\vert \, \Psi_{out}^\psi \big\rangle \big\vert \leq \beta$. Then $Q^\ast$ must make
\begin{align}
\label{eq:alpha-beta}
	\Omega \left(\, \frac{\alpha - \beta}{\sqrt{\gamma}}\,\right)
\end{align}
accesses to a c-$U_\psi$ or c-$U_\phi$ to obtain these output states.
\end{theo}

This theorem can be applied to our case as follows. The input state consists of $c$ copies of $\ket{\ppi}$, while the output state consists of $c+1$ classical samples from $\ppi$. Namely, we are considering a quantum circuit $\mathcal{Q}^\ppi$ which aims at producing these $c+1$ classical samples. $\mathcal{Q}^\ppi$ consists of a sequence of CPTP maps making oracle calls to c-$U_\ppi$, controlled reflections around $\ket{\ppi}$; then, purifying the maps of $\mathcal{Q}^\ast$ to unitary operations we obtain a circuit $Q^\ast$ which employs the same number of oracle calls to c-$U_\ppi$. The output of $Q^\ppi$ then has the form:
\begin{align}
\label{eq:Q}
	Q^{\ppi} \big( \, \ket{\ppi}^{\otimes c} \, \ket{0} \, \big) 
	\; = \; 
	\sum_{x_1, \ldots, x_{c+1}} 
	\sqrt{\pi(x_1) \cdots \pi(x_{c+1})} \, 
	\ket{x_1, \ldots, x_c} \ket{\phi(x_1, \ldots, x_{c+1})} \;,
\end{align}
where $\ket{\phi(x_1, \ldots , x_{c+1})}$ is a state containing all the residual information. The output state in Eq.~\eqref{eq:Q} also generalizes other tasks, \textit{e.g.}, choosing $\ket{\phi(x_1, \ldots , x_{c+1})} = \ket{0}$ corresponds to preparing $c+1$ copies of $\ket{\ppi}$.

Next, we consider a set $\mathcal{Z}$ of quantum states which are coherent encodings of specific probability distributions; on these states we impose a relation $\mathcal{R}$ which is suitable for computing a bound as in Eq.~\eqref{eq:gamma}. In turn, using Eq.~\eqref{eq:alpha-beta}, this will provide a lower bound to the number of reflectors required by the quantum circuit $\mathcal{Q}^\ast$ or, equivalently, by its purification $Q^\ast$.

\begin{prop}
\label{prp:Prop}
We consider the set $\mathcal{Z}$ of states of the form 
\begin{align}
	& \ket{\textbf{u}_S} 
	\; := \;
	\frac{1}{\sqrt{K}} \sum_{x \in S} \ket{x} \,,
\end{align}
where $S \subseteq [N]$ is a subset containing a fixed number $K$ of elements, with $K \leq N$. The state $\ket{\textbf{u}_S}$ corresponds to the coherent encoding of the probability distribution $\textbf{u}_S$. Notice that $f(\textbf{u}_S) \equiv \sum_{x\in S} \frac{1}{\sqrt{K}} = \sqrt{K}$, hence the value of $f(\textbf{u}_S)$ can take any value in the interval $[1, \sqrt{N}]$. Moreover, we say that two states $\ket{\textbf{u}_S},\ket{\textbf{u}_{S'}}$ are in relation $\mathcal{R}$ iff
\begin{align}
	\braket{\textbf{u}_S}{\textbf{u}_{S'}} = a
	\quad \Longleftrightarrow \quad
	 |S \cap S'| = a K \;,
\end{align}
where $a \in (0,1)$ is a constant.

Then, the inequality~\eqref{eq:gamma} can be expressed as follows: $\forall\, S \subseteq [N]$ with $|S| = K$, $\forall\, \ket{\eta}$ orthogonal to $\ket{\textbf{u}_S}$
\begin{align}
\label{eq:gamma'}
	\underset{S':\, |S'| = K \atop |S \cap S'| = aK}
	{\mathds{E}}
	\Big[ \ \big\vert
	\braket{\eta}{\textbf{u}_{S'}}
	\big\vert^2 \; \Big]
	\; \leq \; 
	\frac{a}{K} + 6 \, (1-a)^2 \,\frac{K}{N}
	\; \equiv \; 
	\gamma \;,
\end{align}
provided that $1/(1-a) \leq K \leq N/2$.
\end{prop}

\begin{proof}
First, notice that $\ket{\eta} = \sum_x \eta_x \,\ket{x}$ is orthogonal to $\ket{\textbf{u}_S}$, hence $\sum_{x\in S} \eta_x = 0$. Then we expand:
\begin{align} \label{eq:expectation}
	\underset{S':\, |S'| = K \atop |S \cap S'| = aK}
	{\mathds{E}}
	\Big[ \ \big\vert
	\braket{\eta}{\textbf{u}_{S'}}
	\big\vert^2 \; \Big] \ 
	& = \
	\underset{S':\, |S'| = K \atop |S \cap S'| = aK}
	{\mathds{E}}
	\Big[ \ \Big\vert
	\sum_{i \in S'} \frac{\eta_i}{\sqrt{K}}
	\Big\vert^2 \; \Big] 
	\nonumber\\ 
	& = \
	\frac{1}{K}
	\underset{S':\, |S'| = K \atop |S \cap S'| = aK}
	{\mathds{E}}
	\Big[ \ 
	\sum_{i \in S'} \sum_{j \in S'} \eta_i^\ast \eta_j
	\; \Big]
\end{align}
Next, we split the sum over elements in $S'$ as sum of elements in $S'\cap S$ and elements in $S' \setminus S$:
\begin{align} \label{eq:sum_of_terms}
	\eqref{eq:expectation} 
	\ = \ & \frac{1}{K}
	\underset{S':\, |S'| = K \atop |S \cap S'| = aK}
	{\mathds{E}}
	\Big[ \ 
	\sum_{i \in S'\cap S} \sum_{j \in S'\cap S} 
	\eta_i^\ast \eta_j +
	\sum_{i \in S'\setminus S} \sum_{j \in S'\setminus S}
	\eta_i^\ast \eta_j + 
	\nonumber \\
	& \qquad \qquad \quad \ \
	+ \sum_{i \in S'\cap S} \sum_{j \in S'\setminus S} 
	(\eta_i^\ast \eta_j + \eta_j^\ast \eta_i)
	\; \Big] 	
\end{align}
The first term in the sum in Eq.~\eqref{eq:sum_of_terms} evaluates to
\begin{align}
	\underset{I \subseteq S: \atop |I| = aK}
	{\mathds{E}}
	\Big[ \ 
	\sum_{i \in I} \sum_{j \in I} \eta_i^\ast \eta_j
	\; \Big] \ 
	& = \
	a \sum_{i \in S} |\eta_i|^2
	+ \frac{aK(aK-1)}{K(K-1)} \sum_{i,j \in S \atop i \neq j} \eta_i^\ast \eta_j \;;
\end{align}
the second term in the sum evaluates to ($S^c$ is the complementary of $S$)
\begin{align}
	&\underset{I \subseteq S^c: \atop |I| = (1-a)K}
	{\mathds{E}}
	\Big[ \ 
	\sum_{i \in I} \sum_{j \in I} \eta_i^\ast \eta_j
	\; \Big]
	\ = 
	\frac{(1-a)K}{N-K} \sum_{i \in S^c} |\eta_i|^2
	+ \frac{(1-a)K[(1-a)K-1]}{(N-K)(N-K-1)} \sum_{i,j \in S^c \atop i \neq j} \eta_i^\ast \eta_j \;;
\end{align}
while the third term evaluates to
\begin{align}
	& \underset{S':\, |S'| = K \atop |S \cap S'| = aK}
	{\mathds{E}}
	\Big[ \ 
	\sum_{i \in S' \cap S} \sum_{j \in S'\setminus S} 
	( \eta_i^\ast \eta_j + \eta_j^\ast \eta_i)
	\; \Big]
	\ = \
	\frac{aK}{K} \frac{(1-a)K}{N-K} 
	\sum_{i \in S \atop j \in S^c} (\eta_i^\ast \eta_j + \eta_j^\ast \eta_i) \;.
\end{align}
Finally, using the equation $\sum_{x\in S} \eta_x = 0$, we get:
\begin{align}
	\underset{S':\, |S'| = K \atop |S \cap S'| = aK}
	{\mathds{E}}
	\Big[ \ \big\vert
	\braket{\eta}{\textbf{u}_{S'}}
	\big\vert^2 \; \Big] \
	& = \
	\frac{a}{K} \sum_{i \in S} |\eta_i|^2 + 
	\frac{1-a}{N-K} 
	\bigg(
	\sum_{i \in S^c} |\eta_i|^2
	+ \frac{(1-a)K-1}{N-K-1} \sum_{i,j \in S^c \atop i \neq j}
	\eta_i^\ast \eta_j
	\bigg) \nonumber\\
	& \leq \
	\frac{a}{K} + 
	\frac{1-a}{N-K} 
	\bigg(
	1 + \frac{(1-a)K-1}{N-K-1} (N-K)
	\bigg) \nonumber\\
	& \leq \
	\frac{a}{K} + 
	\frac{1-a}{N-K} \bigg(
	1 + 2 \, (1-a)K
	\bigg)\nonumber\\
	& \leq \
	\frac{a}{K} + 
	\frac{1-a}{N-K}
	\, 3 \, (1-a)K
	\nonumber\\
	& \leq \
	\frac{a}{K} + 
	6 \, (1-a)^2 \, \frac{K}{N}
\end{align}
for appropriate choices of $K$; namely, we have used respectively $K \leq N-2$, $K \geq 1/(1-a)$ and $K \leq N/2$ for the last three inequalities in the derivation above. 
\end{proof}

\begin{coro}
Consider the family of quantum states $\mathcal{Z}$ and the relation $\mathcal{R}$ given in \prp{Prop}, setting the constant $a = 1 - 1/c$. Consider then a (purified) quantum circuit $Q^\ast$ having access to controlled reflections around $\ket{\textbf{u}_S}$. The input states to $Q^\ast$ have the form $\big\vert\, \Psi_{in}^\psi \,\big\rangle = \ket{\textbf{u}_S}^{\otimes c}$, $\big\vert\, \Psi_{in}^\phi \,\big\rangle = \ket{\textbf{u}_{S'}}^{\otimes c}$. The output states $\big\vert\, \Psi_{out}^\psi \,\big\rangle$ and $\big\vert\, \Psi_{out}^\psi \,\big\rangle$ have the same form as the right hand side of Eq.~\eqref{eq:Q} for $\ppi = \textbf{u}_S$ and $\ppi = \textbf{u}_{S'}$, respectively. 

Then the circuit $Q^\ast$ makes $\frac{1}{c^2} \Omega	 \big(\, C(\textbf{u}_S) \,\big)$ controlled reflections around $\ket{\textbf{u}_S}$.
\end{coro}

\begin{proof}

First, substituting $a = 1 - 1/c$ in Eq.~\eqref{eq:gamma'} we obtain:
\begin{align}
	\underset{S':\, |S'| = K \atop |S \cap S'| = K - K/c}
	{\mathds{E}}
	\Big[ \ \big\vert
	\braket{\eta}{\textbf{u}_{S'}}
	\big\vert^2 \; \Big]
	\; \leq \; 
	\frac{1}{K} + 6 \, c^2 \,\frac{K}{N}
	\; \equiv \; 
	\gamma' \;.
\end{align}
Second, notice that for our choice of input and output states:
\begin{align}
	\big\vert \big\langle \, \Psi_{in}^\phi \, 
	\big\vert \, \Psi_{in}^\psi \big\rangle \big\vert 
	\; & = \; a^c \\
	\big\vert \big\langle \, \Psi_{out}^\phi \, 
	\big\vert \, \Psi_{out}^\psi \big\rangle \big\vert 
	\; & \leq \; a^{c+1} \;.
\end{align}
Then, the application of \thm{Aaronson} for $\alpha = a^c$ and $\beta = a^{c+1}$ directly yields a $\Omega \left(\, \frac{a^c - a^{c+1}}{\sqrt{\gamma'}}\,\right)$ lower bound to the number of oracle access that are required by $Q^\ast$. Setting $a = 1 - 1/c$ this lower bound becomes $\Omega\left( \frac{1} { c\, \sqrt{\gamma'} }\right)$ and, finally, this can be further simplified to obtain the required result:
\begin{align}
	\Omega
	\left(
	\frac{1}{c \, \sqrt{\gamma'}}
	\right) 
	\ & \geq \
	\Omega
	\left(
	\min 
	\left\lbrace 
	\frac{1}{c^2} \sqrt{\frac{N}{K}} \; , \;
	\frac{1}{c} \sqrt{K}
	\right\rbrace
	\right)
	\nonumber\\
	\ & \geq \
	\frac{1}{c^2} \
	\Omega
	\left(
	\min 
	\left\lbrace 
	\frac{\sqrt{N}}{f(\textbf{u}_S)} \; , \;
	f(\textbf{u}_S)
	\right\rbrace
	\right)
	\nonumber\\
	\ & = \
	\frac{1}{c^2} \
	\Omega
	\big(\,
	C(\textbf{u}_S)
	\,\big) \;.
\end{align}

\end{proof}

This Corollary shows that in general $\O(C(\ppi))$ reflections around $\ket{\ppi}$ are needed in order to obtain multiple samples from $\ppi$. Using the results of Corollary 5.3 and 5.4 of~\cite{2013_Aaronson} the same $\O(C(\ppi))$ lower bound applies for drawing approximate samples from $\ppi$, say, within constant approximation error. This is important since the distribution $\textbf{u}_S$ cannot be, strictly speaking, a stationary distribution of a irreducible MC, since it would need to have support over the entire set of $N$ elements.

\end{document}